\documentclass[12pt]{article}

\usepackage{amsbsy} \usepackage{amssymb} \usepackage{amsmath}
\makeatletter
\renewcommand\section{\@startsection {section}{1}{\z@}%
                                 {-3.5ex \@plus -1ex \@minus -.2ex}
                                   {2.3ex \@plus.2ex}%
                                   {\normalfont\large\bfseries}}
\renewcommand\subsection{\@startsection{subsection}{2}{\z@}%
                                   {-3.25ex\@plus -1ex \@minus -.2ex}%
                                     {1.5ex \@plus .2ex}%
                                     {\normalfont\bfseries}}
\renewcommand\subsubsection{\@startsection{subsubsection}{3}{\z@}%
                                   {-3.25ex\@plus -1ex \@minus -.2ex}%
                                     {1.5ex \@plus .2ex}%
                                     {\normalfont\itshape}}
\makeatother

\newcommand{\non}{\nonumber \\} 
\newcommand{\noi}{\noindent}

\def\alp{\alpha} 
\def\bet{\beta} 
\def\gam{\gamma} 
\def\del{\delta}
\def\eps{\epsilon} 
 
\def\zet{\zeta}

\def\lam{\lambda} 
 
\def\sig{\sigma}

\def\Gam{\Gamma}

\def\Sig{\Sigma}

\def\cD{{\cal D}} 
 \def\cF{{\cal F}} 
   
  \def\cM{{\cal M}} 
\def\cO{{\cal O}} \def\cP{{\cal P}}

\def\pa{\partial}

\def\rar{\rightarrow}  

\def\one{1\!\!1\,\,}

\newcommand{\tr}{\mbox{Tr}}

\def\sym{{\rm Sym}} 
\def\mtrix#1{\begin{matrix} #1
\end{matrix}} 
\newcommand{\str}{\mathrm{Str}}

\begin{document}

\thispagestyle{empty}
\begin{flushright}
\parbox[t]{2in}{CU-TP-1100\\ 
ITFA-2003-50\\ 
hep-th/0310150}
\end{flushright}

\vspace*{0.2in}

\begin{center}
{\Large \bf General covariance of the non-abelian DBI-action:
\\[0.1in] Checks and Balances }

\vspace*{0.3in} {Jan de Boer${}^{1}$, Koenraad
Schalm${}^{2}$~ and Jeroen Wijnhout${}^{1}$}\\[.3in]

${}^1${\em Institute for Theoretical Physics \\ 
University of Amsterdam \\ 
Valckenierstraat 65\\ 
1018XE Amsterdam }\\[.1in] 
${}^2${\em Department of Physics \\ 
Columbia University \\ 
New York, NY 10027}\\[.2in]

{\bf Abstract}
\end{center}
We perform three tests on our proposal to implement diffeomorphism
invariance in the non-abelian D0-brane DBI action as a {\em basepoint
  independence} constraint between matrix Riemann normal coordinate
systems. First we show that T-duality along an isometry correctly
interchanges the potential and kinetic terms in the action. Second, we
show that
the method to impose basepoint independence using an auxiliary
$dN^2$-dimensional non-linear sigma model also works for 
metrics which are
curved along the brane, provided a physical gauge choice is made at
the end. 
Third, we show that without alteration this method is applicable
to higher order in velocities. Testing specifically to order four, we
elucidate the range of validity of the {symmetrized}  trace
approximation to the non-abelian
DBI action.

\newpage

\section{Introduction}

The nature of spacetime at the smallest scales is still an open
question, many recent advances in non-perturbative 
string theory notwithstanding. Yet, the fundamental principle 
that the laws of physics should be observer independent, 
leads us to expect that even at the smallest scales 
general coordinate invariance manifests itself in some form
or other. The one explicit
proposal for non-perturbative string theory we currently possess, 
M(atrix) theory, contains explicit couplings 
to the graviton. M(atrix) theory is
formulated in the light-cone gauge, and not background independent,
but these couplings ought to 
reflect the freedom to choose different (transverse) coordinates. This
expectation is based on the connection between M(atrix)
theory and the low-energy-effective-action for $N$ superposed
D0-branes. The latter is derivable from string theory and constitutes a
generalization of the invariant length 
point particle action $S= \int ds = \int
\sqrt{g_{ij}\dot{x}^i\dot{x}^j}$ where the coordinates $x^{\mu}$
have been promoted to $U(N)$-valued non-abelian matrices $X^{\mu,ab}$,
together with
a potential term proportional to commutators. 
Such matrix-valued coordinates no longer commute, and it begs
the question how to implement the nonlinear coupling to
gravity. Gravity is the gauge theory which imposes general coordinate
invariance, and the problem is therefore equivalent to finding the
invariant 
action for non-abelian D0-branes in a curved background.

Guided by these symmetry considerations 
we attempted in \cite{DeBoer:2001uk} to construct the full coupling of
$N$ non-abelian D0-branes to gravity by imposing that the action be
invariant under general coordinate transformations. Naturally, the
matrix-valued nature of the coordinates $X^{\mu}$ goes beyond
Riemannian 
geometry and 
general coordinate transformations and the diffeomorphism group will
take a wholly new form. Moreover, the non-abelian nature of the
coordinates does 
make the problem a very difficult one. Considering maps between
(matrix) 
Riemann normal coordinate systems centered on different base points
$\cP_i$, we put forth that diffeomorphism invariance can be
implemented as a new symmetry principle: {\em base-point
independence}. The advantage of this method is that (i) the definition
of matrix normal coordinates --- that $X^{i}(\tau)=\tau Y^i$ with
$Y^i$ constant is a solution to the non-abelian geodesic equation,
which is the {\em field equation} of the action $S$ ---
yields an additional set of constraints on the action, and (ii) in
normal 
coordinates all
nonlinear terms in the action are tensors at the basepoint $\cP$ and
transform {\em 
  covariantly} under change of basepoint. These two points plus 
the constraint that the action itself should be invariant under a
basepoint transformation allowed us to construct an algorithm to
determine the action order by order in the matrices $X^{\mu}$. Solving
the algorithm explicitly to first nontrivial order $\cO(X^6)$, we
found a number of surprises:
\begin{itemize}
\item[(i)] In contrast to the abelian point particle action, the
 more stringent constraints imposed by base-point independence do {\em
 not} determine the curved space action for non-abelian D0-branes 
 uniquely. Signs that this would be the case
 had been found earlier: at the linearized level in diffeomorphisms,
 two different stress tensors are compatible with current conservation
 \cite{Okawa:2001if}. One arises in the low-energy 
effective action (LEEA) for
 non-abelian D0-branes in the bosonic string; the second appears in
 the  LEEA for non-abelian D0-branes in type II superstring theory.   
\item[(ii)] Despite the remaining arbitrariness in the action, 
  one can show that (a) a fully symmetrized trace
  structure at all orders is incompatible with base-point independence
  and (b) compared to the abelian case, there are always new
  vertices, i.e. couplings to the gravitational background
  proportional to commutators. 
\end{itemize}
A consequence of (ii) is that the gravitational potential felt by
non-abelian D0-branes is fundamentally different than that felt by
abelian particles. Exemplifying this point is the evidence that in
hyperbolic spacetimes such
non-abelian D0-branes behave collectively rather than independently:
there are signs that a gravitational analogue to the Myers effect
exists \cite{Myers,deBoer:2002we}. The order $\cO(X^6)$ base-point
independent action furthermore 
obeys all Douglas's axioms of D-geometry: necessary
properties of the action for non-abelian D0-branes in curved space
\cite{Douglas:1997ch}. In particular, with a specific ansatz for a
generalization of the flat-space potential to one built out of
manifestly base-point independent objects (see section
\ref{sec:checks:-t-duality}) the masses of fluctuations around a
diagonal configuration are given by the geodesic distances between the
diagonal entries.

Complementary to these results is recent work of van Raamsdonk
on gauge-theories on
``spaces'' 
with $U(N)$-valued coordinates \cite{vRaamsdonk}, which confirmed a
number of qualitative aspects. 
In string theory, these theories arise
as the LEEA of D0-branes embedded in the worldvolume of a
higher-dimensional D$p$-brane. String theory similarly predicts that 
a LEEA of non-abelian D0-branes
coupled to gravity should exist. Yet with the strong constraints
imposed by base-point independence for {\em matrix-valued}
coordinates, it is a wonder we were able to find a solution at all,
let alone a family of solutions. To support the answer we found in
\cite{DeBoer:2001uk} and confirm the consistency of the solutions, we
perform here three tests on our answer and the underlying idea of
base-point independence. In section
\ref{sec:checks:-t-duality} we will show that T-dualizing along an
isometry transforms the ansatz made for the potential term in
\cite{DeBoer:2001uk} into the base-point independent kinetic
term. This confirms our ansatz for the potential. A
shortcoming of the base-point independence method as put forth in
\cite{DeBoer:2001uk} we correct in section
\ref{sec:checks:-non-transv}. In \cite{DeBoer:2001uk} we only
addressed metric spaces with curvature strictly transverse to the the
D0-brane worldvolume. Here we show that our method extends to spaces
with curvature in all directions. In doing so, we solve the paradox
between coordinate invariance and the distinct nature between
tangential (commuting) and transverse (non-abelian) coordinates with
respect to the worldvolume. The resolution lies in an extended
definition of the 'physical gauge'. Finally in section
\ref{sec:checks:-fourth-order} we extend the method of base-point
independence to higher derivative terms in LEEA of non-abelian
D0-branes. This is a necessary condition for our philosophy
to be consistent, and it elucidates the range of
applicability of the symmetrized trace approximation for non-abelian
D0-branes. We will find in particular that the symmetrized trace prescription
is incorrect for terms linear in the graviton and of fourth order in
velocities.
We begin, however, with a brief review of diffeomorphism
invariance for matrix-valued coordinates and base-point independence.
  
\section{Diffeomorphisms, covariant background field formalism and\\ base-point
  dependence} 
\setcounter{equation}{0}
\label{sec:diff-covar-bg}

\subsection{The base-point transformation}

A guiding rule in theoretical physics is to keep as many symmetries
manifest as possible. In an action describing fluctuations, this rule
is automatic for a linearly
acting symmetry: in that case both the background and the fluctuation
transform in the same way. When a symmetry acts nonlinearly on fields,
this rule is a priori difficult to keep. Consider general
coordinate invariance for a first quantized particle action, $S= \int
\sqrt{g_{ij}\dot{x}^i\dot{x}^j}$. The explicit reason why the symmetry
is difficult to maintain, is that the quantum fluctuation $\del
x^i(\tau)= x^{i}(\tau)-x_{bg}^i(\tau)$ is not a covariant object under
background coordinate transformations $\del x_{bg}^i =
\eps^i(x_{bg})$. The resolution is to expand the fluctuations
nonlinearly as well, in such a way that they become covariant. Let
$\xi^i(\tau)$ be the tangent vector $\xi^i(\tau)$ at $x_{bg}^i$ along
the geodesic towards $x^i(\tau)$, and solve for the geodesic between
$x^i(\tau)$ and $x^i_{bg}(\tau)$ in 
terms of $\xi^i$:
\begin{equation}
\label{eq:1}    
x^i = x_{bg}^i + \xi^i - \sum_{n=2}^\infty \frac{1}{n!}\Gamma^i_{~j_1
    \ldots j_n}(x_{bg}) \xi^{j_1}\cdots\xi^{j_n}.
\end{equation}
Here $\Gamma^i_{~j_1\ldots j_n} \equiv \nabla^{cov}_{(j_1}
\Gamma^i_{~~j_2\ldots j_n)}$ are generalized connection 
symbols, where $\nabla^{cov}_{j_1}$ only acts on the lower indices;
see for example \cite{Alvarez-Gaume:hn,mukhi}. As $\xi^i$ is a vector,
it transforms co(ntra)variantly under background gauge
transformations. Eq. \eqref{eq:1} therefore 
constitutes a (nonlinear) expansion in
fluctuations $\xi^i$ which is consistent with the symmetries.

We can also {\em use} the symmetry to our advantage. Eq. \eqref{eq:1}
also defines a coordinate transformation between coordinates
$x^i$ and (Riemann normal) coordinates $\xi^i$. In this
new coordinate system $x^i_{bg}$ is the origin and it is not difficult
to show that all geodesics through $x^i_{bg}$ are straight lines. In normal
coordinates $\xi^i$, the
geodesic connecting $\xi^i$ and the origin $x^i_{bg}$ reads 
\begin{eqnarray}
  \label{eq:2}
  \xi^i(\tau) = \tau \xi^i~.
\end{eqnarray}
As eq. \eqref{eq:1} holds for all coordinate systems, this in turn
implies that in Riemann 
normal coordinates (RNC) the generalized connection
coefficients vanish at the origin
\begin{eqnarray}
  \label{eq:3}
  \Gamma^i_{~j_1\ldots j_n}(x^i_{bg})|_{{\rm RNC~around}~x^i_{bg}} =
  0~.
\end{eqnarray}
The action for the fluctuations, built out of objects evaluated at
$x^i_{bg}$ therefore contains no connection terms. It consists only of
true tensors at $x^i_{bg}$ and the coordinate $\xi^i$, originally a
tangent vector at $x^i_{bg}$ and will be manifestly covariant under
background coordinate transformations. 
\begin{displaymath}
    S^{RNC}_{exp} = S[g(x_{bg}),R(x_{bg}),\nabla R(x_{bg}),\ldots,\xi].
\end{displaymath}
In fact, a moment's thought
reveals that because the fluctuation 
$\xi$ transforms co(ntra)variantly, the action after the expansion
\eqref{eq:1} in any coordinate system is built from covariant
quantities. We will see below, though, that the specific choice of normal
coordinates is very helpful.

This non-linear expansion is familiar as the non-linear sigma model
version of the covariant background field expansion
\cite{Alvarez-Gaume:hn,mukhi}. To discuss what {\em base-point
  independence} means, it is useful to recall the covariant BG field
method for Yang-Mills theories. The
covariant BG field method entails a split of the gauge connection
$A_{\mu} = A^{bg}_{\mu}+Q_{\mu}$ into a background part $A_{\mu}^{bg}$
and a quantum fluctuation $Q_{\mu}$. The gauge
transformation 
\begin{eqnarray}
  \label{eq:4}
  \del A_{\mu} = \del (A^{bg}_{\mu} +Q_{\mu}) =
  D_{\mu}(A^{bg}+Q)\Lambda 
\end{eqnarray}
decomposes into a standard gauge transformation of the background
field, 
\begin{eqnarray}
  \label{eq:5}
  \del_{bg} A_{\mu}^{bg} =
  D_{\mu}(A^{bg})\Lambda~, 
\end{eqnarray}
plus a covariant gauge rotation of the quantum field $Q_{\mu}$,
\begin{eqnarray}
  \label{eq:6}
  \del_{bg} Q_{\mu} = [Q_{\mu},\Lambda]~.
\end{eqnarray}
Essential for proving equivalence with the standard approach to
correlation functions, is that the background expanded action only
depends on the combination $A_{\mu}^{bg}+Q_{\mu}$: The background
expanded action has an additional shift symmetry $A^{bg}_{\mu} \rar
A^{bg}_{\mu}+\eps_{\mu};~Q_{\mu} \rar Q_{\mu} -\eps_{\mu}$
\cite{Abbott:zw}. Vice
versa, suppose an action is invariant under both this shift symmetry
and the background gauge transformations \eqref{eq:5} and
\eqref{eq:6}. The true symmetry of the action is then
eq. \eqref{eq:4}, and we recover the standard YM action.

Compare this with the background field method for the
non-linear sigma model. Analogous to
eqs. (\ref{eq:4})-(\ref{eq:6}) the now 
non-linear expansion in fluctuations \eqref{eq:1} guarantees that the
orginal general coordinate invariance 
\begin{eqnarray}
  \label{eq:7}
  \del x^i = \del (x^i_{bg}+ \xi^i-''\Gam'') = \eps^i(x= x_{bg}+ \xi-''\Gam'')
\end{eqnarray}
decomposes into a standard
coordinate transformation for the background field,
\begin{eqnarray}
  \label{eq:8}
  \del_{bg} x_{bg}^i = \eps^i(x_{bg})~,
\end{eqnarray}
plus a co(ntra)variant transformation of the quantum fluctuation,
$\xi^i$
\begin{eqnarray}
  \label{eq:9}
  \del_{bg} \xi^i = -\eps^j(x_{bg}) \pa_{j} \xi^i + \xi^k \pa^i
  \eps_k(x_{bg})~. 
\end{eqnarray}
In addition the background expanded action should have a ``shift''
symmetry which guarantees that it only depends on the particular
combination of $x^i_{bg}$ and $\xi^i$ in \eqref{eq:1}.  Due to the nonlinear
nature of the expansion, this ``shift'' symmetry will be nonlinear as
well. To see what this
``shift'' symmetry exactly is, we hark back 
the geometric principles underlying the background field
expansion. In particular, recall that the normal coordinate system is
defined by its properties with respect to the origin, i.e. the point
$x_{bg}$. This suggest that we can
compensate for a shift symmetry $x_{bg}^i \rar
x_{bg}^i
+\eps^i(x_{bg})$ by choosing a new RNC coordinate system around a new
'basepoint' $\tilde{x}_{bg}\equiv x_{bg}^i
+\eps^i(x_{bg})$. By construction this compensating coordinate
transformation to a new set of RNC is given by considering the
geodesic from $\tilde{x}_{bg}$ to $\xi$ \footnote{In the
  last step we have used that in the old RNC the connection
  coefficients 
vanish at the origin, 
viz. 
$$
\Gam^i_{~jk}(\tilde{x}_{bg})\simeq \eps^l\pa_l\Gam^i_{~jk}({x}_{bg}) \stackrel{RNC}{=}
-\frac{1}{3}\eps^l R_{l(jk)}^{~~~~~i}(x_{bg}) ~.
$$
This also sets our convention for the Riemann tensor.}
\begin{eqnarray}
  \label{eq:10}
  &&x_{bg}+ \xi^i = \tilde{x}_{bg}+\chi- \sum_{n=2}^\infty \frac{1}{n!}\Gamma^i_{~j_1
    \ldots j_n}(\tilde{x}_{bg}) \chi^{j_1}\cdots\chi^{j_n}.\\ 
&\Rightarrow & \xi^i = \epsilon^i + \chi^i + \eps^k
    \sum_{n=2}^\infty\frac{1}{(n+1)!}
    \nabla_{j_n}\ldots\nabla_{j_3}R_{j_1 (ki)j_2}(0)\chi^{j_1} \cdots
    \chi^{j_n}.
    \label{eq:bpt}
\end{eqnarray}
The tangent vector $\chi^i$ at $\tilde{x}_{bg}$ 
from $\tilde{x}_{bg}^i$ to $\xi^i$ is the new normal coordinate around
$\tilde{x}_{bg}$. 
Furthermore, due to the special properties of RNC, all tensors --- the other
building blocks of the action --- transform {\em covariantly} under this
``shift''. Schematically
\begin{eqnarray}
  \label{eq:11}
   R(x_{bg}) \rightarrow R(\tilde{x}_{bg}) = R(x_{bg})+ \epsilon\pa
    R(x_{bg}) \stackrel{RNC}{=} R(x_{bg}) + 
    \epsilon\nabla R(x_{bg}).
\end{eqnarray}
Thus if the action is invariant under this ``shift'' of basepoint,
i.e.
\begin{eqnarray}
    S &=& S[g(x_{bg}),R(x_{bg}),\ldots,\epsilon+\chi
    + \eps R(x_{bg}) \chi \ldots \chi]\non
  & =& S[g(x_{bg}) + \epsilon \nabla g(x_{bg}),R(x_{bg})+\epsilon\nabla
    R(x_{bg}),\ldots,\chi]\non
& =& S[g(\tilde{x}_{bg}),R(\tilde{x}_{bg}),\ldots,\chi],
\end{eqnarray} 
it in fact only depends on the nonlinear combination
\eqref{eq:1}.\footnote{From the technical perspective
  on the covariant background field expansion, the choice of
  RNC corresponds to a choice of background field which obeys the
  equations of motion. For abelian geometry, 
this is not necessary but it does make the
  geometrical picture clearer.} In
combination with the manifest invariance under background coordinate
transformations, this establishes that the action $S$ is in fact
diffeomorphism invariant. Geometrically the meaning of the ``shift''
symmetry is clear. A point $x$ on a manifold $\cM$ can be reached 
either by
succesive infinitesimal translations along a vector $\xi$ from the
basepoint $x_{bg}$ or by translations along $\chi$ from
$\tilde{x}_{bg}$. General coordinate invariance means that formal
local expressions, e.g. the line element, constructed as a function of
$\xi$ in relation to $x_{bg}$ do not depend on which basepoint one
picks. This is the manifestion of diffeomorphism invariance as ''base-point
independence''.
 
\subsection{Generalization to matrix geometry}
 The above summary establishes why general coordinate invariance is
 equivalent to a {\em base-point independent} action for covariant
 fluctuations with tensorial 
 couplings. Importantly, string theory tells us  
 that it is precisely the
 vector-like fluctuations of the
 action for non-abelian D0-branes which become matrix-valued. It is
 therefore 
 more natural to impose general coordinate invariance in the guise of
 base-point independence rather than a matrix generalization of
 Riemannian geometry. We should caution that, though certainly a
 necessary condition, base-point independence may not be truly
 equivalent to a ``diffeomorphism'' invariance for matrix-valued
 coordinates.\footnote{For instance, it is unclear if the base-point
 independence constraint 
also correctly accounts for 
``coordinate transformations'' strictly proportional to
 commutators $\del X \sim [X,X]$. At the same time, from the string
 theory point of view, it is unclear to
 what extent these are really geometrical (see e.g. 
\cite{Hassan:2003uq}).}   
We simply do not know enough about the latter, and we
 will proceed on the assumption that it is so.

With this input from string theory that it is the vectorlike
fluctuations which are promoted to matrices, and that we should
therefore impose diffemorphism invariance as base-point indepence, 
the way to construct a/constraints on the LEEA for non-abelian
D0-branes are clear.
\begin{itemize}
\item [(i)] Write the most general two-derivative action in
  $U(N)$-valued fluctuations $X^{i,ab}$, invariant under $U(N)$
  rotations (it should be a single trace, see below), and with
  tensorial couplings evaluated at an {\em abelian} basepoint
  $x_{bg}$; the origin of the normal coordinate system.
\item[(ii)] Enforce that the action is indeed in 
 matrix normal coordinates: i.e. tune the couplings such that 
 \begin{eqnarray}
   \label{eq:12}
   X^{i,ab}(\tau) = \tau Y^{i,ab}~,~~~~~
 \end{eqnarray}
with $Y^{i,ab}$ a constant matrix, is a solution to the field
equations. In matrix geometry this step is crucial, for it also fixes
novel matrix-type diffeomorphisms of the form (see
e.g. \cite{Hassan:2003uq}) 
\begin{eqnarray}
  \label{eq:13}
  \del X \sim [X,X]~.
\end{eqnarray}
\item[(iii)] Require that the action is (abelian) basepoint
  independent: i.e. solve the field equation 
for {\em matrix-geodesics} between an abelian point
  $\tilde{X}_{bg}^{i,ab}=\epsilon^i\delta^{ab}$ and a matrix point
  $X^{j,cd}$ in terms of the tangent vector
  $Z^{i,ab}=\dot{X}^{i,ab}$. Substitute this coordinate change in the
  action, together with a shift of the background tensors and demand
  that the action be invariant. 
\end{itemize}
We should note that string theory only predicts that coordinates
{\em transverse} to the worldline of the non-abelian D0-brane are
promoted to matrices.\footnote{We have also chosen the gauge $A_0=0$ on
  the worldline. There is a corresponding Gauss's law constraint on
  the matrix-valued coordinates. For the purposes of this paper, it
  will play no role, and we will ignore its consequences.}
Strictly speaking, we therefore consider only
spacetimes which are curved orthogonal to the brane. One of the
purposes of this article is to remedy this situation; we will do so
in section \ref{sec:checks:-non-transv}.

Predictions from string theory place two more constraints on
 the base-point independent
action of non-abelian D0-branes:
 \begin{itemize}
 \item[(iv)] Tseytlin observed that the action must consist of only a
 single trace over $U(N)$-indices \cite{Tseytlin:1997cs}.
\item[(v)] Explicit computations in type II superstring theory 
  have revealed that in the linearized
  weak field approximation (i.e. linear in the small fluctuation
  $g_{\mu\nu}=\eta_{\mu\nu}+h_{\mu\nu}$; in normal coordinates linear
  in the 
  Riemann tensor and its symmetrized derivatives) the ordering is
  completely symmetrized
  \cite{taylorvanraamsdonk,Okawa:2001if}. Surprisingly the symmetrized
  ordering does not arise for D0-branes in the bosonic string. We must
  therefore {\em choose} which representation of matrix-valued
  diffeomorphisms we are interested in, and insist that at the
  linearized level our answer reproduces this. This will partly but
  by no means completely fix the freedom in the action that remains
  after we impose the requirement of diffeomorphism invariance. 
 \end{itemize}
The remaining axioms of D-geometry \cite{Douglas:1997ch} can be shown
to follow from these two constraints plus base-point independence
\cite{DeBoer:2001uk}.  

\subsection{Applying base-point independence}
\label{sec:applying-base-point}
\setcounter{equation}{0}

As formulated these conditions prescribe a consistent algorithm to
find a base-point independent action for matrix-valued normal
coordinates. This direct approach is cumbersome, however, and in
\cite{DeBoer:2001uk} we used a more convenient and intuitive
method. The basis of this ``matrix-geometry'' is the realization
that the final action must be a constrained form of a
$dN^2$-dimensional non-linear sigma model (NL$\sig$M). 
\begin{eqnarray}
  \label{eq:14}
  S= \int d\tau \, G_{IJ}(X) \dot{X}^{I}\dot{X}^J~.
\end{eqnarray}
Each index $I$ describes a triplet $I=\{i;ab\}$ built from a
$d$-dimensional space-time index $i$ and two $U(N)$-indices $a,b$.
The $dN^2$-dimensional metric $G_{IJ}(g_{ij},R_{ijkl},\ldots,X)$ is a
{\em functional} of the $d$-dimensional metric $g_{ij}$ and its
derivatives. Using a $dN^2$-dimensional expansion in normal
coordinates, one must impose the following conditions to obtain the
$d$-dimensional base-point independent action for matrix normal
coordinates:
\begin{itemize}
\item[(a)] When functionally 
expressed in terms of the $d$-dimensional constituents the $dN^2$
dimensional metric, Riemann tensor, and covariant derivatives thereof must obey
{\em all} 
the usual identities of symmetry/antisymmetry, Bianchi identity,
commutation relations of covariant derivatives, etc.
\item[(b)] The $U(N)$ indices must be such that the action is a
  single trace; i.e. no traces may occur within the functional
  expressions.
\item[(c)] It should have the right $U(1)$ limit for diagonal matrices.
\item[(d)] At linearized order the symmetrized ordering should emerge.
\item[(e)] Most importantly, {\em base-point independence} follows
  from the requirement that the ``trace'' of the $dN^2$-dimensional covariant
  derivative, acts as the $d$-dimensional covariant derivative:
  \begin{eqnarray}
    \label{eq:16}
    \delta^{ab} \nabla_{i;ab}({\rm anything}) = \nabla_i({\rm
    anything})~.
  \end{eqnarray}
\end{itemize}
For instance at order 4 and 5 in matrix
normal coordinates $X^{i;ab}$ the two relevant $dN^2$-dimensional
tensors are the Riemann tensor and its covariant derivative. Imposing
the above matrix-geometry constraints one finds that in terms of
$d$-dimensional curvature tensors, they are
\begin{eqnarray}
  \label{eq:17}
  R_{IJKL} &=& R_{ijkl} \Sig_{a_ib_ia_jb_ja_kb_ka_lb_l}~, \non
 \nabla_{M}R_{IJKL} &=& \nabla_{m}R_{ijkl}
 \Sig_{a_mb_ma_ib_ia_jb_ja_kb_ka_lb_l}~, 
\end{eqnarray}
where $\Sigma_{a_1b_1\ldots a_nb_n}$ is the object that when
contracted with $n$-matrices returns the symmetrized trace
\begin{displaymath}
    \Sigma_{a_1b_1,\ldots,a_nb_n} O^{i_1a_1b_1}\cdots O^{i_na_nb_n} =
    \mathrm{Str}(O^{i_1}\cdots O^{i_n})~.
\end{displaymath}
At order six, the fully symmetrized ordering is no longer 
consistent with the identity
$$
[\nabla_{N},\nabla_{M}]R_{IJKL} =
R_{NMI}^{~~~~~\,P}R_{PJKL}+\ldots
$$ 
This illustrates why  
the symmetrized approximation corresponds to the linearized
approximation.

Finally, it will also turn out to be convenient to introduce a
$dN^2$-vielbein $E^A_I$, which we define below, and which will
be used to define the potential and discuss T-duality in section~3.

\subsubsection{Second order in $\dot X$}

To explicitly 
show how the matrix-geometry generates a base-point independent
action, we review here the application for the kinetic term ---
order two in derivatives--- to order $\cO(X^4)$, 
i.e. we show that the action
\begin{equation}
    L_2 = -\tfrac{1}{2} (\delta_{ij}\mathrm{tr}({\dot X}^i{\dot X}^j)
    + \tfrac{1}{3}\mathrm{R}_{iklj}\mathrm{Str}({\dot X}^i{\dot
    X}^jX^kX^l))+\cO(X^5)
    \label{eq:secondorder}
\end{equation}
is base-point independent.

Writing the action in terms of a $dN^2$ NL$\sig$M, we find
\begin{equation}
    L = -\tfrac{1}{2} \eta_{AB} \Pi^A\Pi^B,
    \label{eq:secondordermg}
\end{equation}
with
\begin{displaymath}
    \Pi^A = E^A_I {\dot X}^I.
\end{displaymath}
Matching with the flatspace result, the 
tangent space metric acts as a trace on the matrix indices:
\begin{equation}
\label{eq:15}
    \eta_{AB} = \eta_{iab,jcd} = \delta_{ij}
    \delta_{ad}\delta_{bc}.
\end{equation}
The metric $\eta_{AB}$ is a twisted version of the $SO(dN^2)$ invariant
metric; 'twisted' means that it equals a Wick-rotated $SO(dN^2)$
metric up to a similarity transformation.\footnote{E.g. for $N=2$, one
  finds
$$
\eta_{iab,jcd}=\delta_{ij}\left(\begin{matrix} 1 & 0 & 0 & 0 \cr
                                         0 & 0 & 1 & 0 \cr
                                         0 & 1 & 0 & 0 \cr
                                         0 & 0 & 0 & 1 \end{matrix}\right)
$$
with rows (columns) labeled by $ab=\{11,12,21,22\}$. Hence in the
$N=2$ case the 
twisted metric is that of $SO(3d,d)$.} Note that this is slightly
different from the convention used in \cite{DeBoer:2001uk}.

Expanding in RNC the vielbein equals:
\begin{equation}
    E^A_I = \delta^A_I + \frac{1}{12} R^A_{\phantom{A}(PQ)I} X^PX^Q +
    \tfrac{1}{24} \nabla_P R^A_{\phantom{A}(QR)I} X^PX^QX^R + \ldots
\end{equation}
Substituting this 
into equation \ref{eq:secondordermg}, and using \eqref{eq:17}, 
we recover equation \eqref{eq:secondorder}. By virtue of the fact that
eq. \eqref{eq:17} is the solution to the
matrix-geometry constraints, this action is basepoint independent. For
this simple case, one can check it explicitly \cite{DeBoer:2001uk}.

An instructive illustration of the power of the matrix-geometry
method, is the following exercise.
Although we know from the flat space limit that the one-form $\Pi^A$
should be contracted with the tangent space metric $\eta_{AB}$ of
eq. \eqref{eq:15}, we can consider a more general case.  
\begin{equation}
    L_2 = -\tfrac{1}{2}M_{AB}(X) \Pi^A \Pi^B,
\end{equation}
Expanding in RNC as prescribed one obtains the action 
(with the base-point $\bar X$, see \cite{mukhi} for a
convenient algorithm):
\begin{equation}
\begin{split}
    &-2 L_2 \\ &= \left \{ \left . M_{AB} \right |_{\bar X} + \left
. \nabla_C M_{AB} \right |_{\bar X} X^C + \tfrac{1}{2} \left
. (\tfrac{2}{3} M_{QB} \mathrm{R}^Q_{\phantom{Q}CDA} +
\nabla_D\nabla_C M_{AB} )\right |_{\bar X} X^C X^D +\ldots \right \} {\dot
X}^A{\dot X}^B.
\end{split}
\end{equation}
Comparing with the flat space case we read off: $\left . M_{AB}
\right |_{\bar X} = \eta_{AB}$. Assume that it is
possible to set $\nabla \ldots \nabla \left . M \right |_{\bar X} =
0$, then we get $M(X)_{AB} = \eta_{AB}$ (remember: in RNC partial and
covariant 
derivates are the same). This results in the action:
\begin{equation}
\begin{split}
    -2L_2 &= (\eta_{AB} + \tfrac{1}{6} \mathrm{R}_{B(CD)A} X^CX^D)
    {\dot X}^A{\dot X}^B \\ &= \eta_{AB} \Pi^A\Pi^B.
\end{split}
\end{equation}
Given that the vielbein is constructed from tensors obeying the
matrix-geometry constraints, 
we can check the properties the action needs to
have, without explicit calculations. We only need to verify that
$M_{AB}$ also satisfies the matrix-geometry constraints. 
Single traceness of the action follows from the fact that 
$M_{AB}=\eta_{AB}$ 
has no internal $U(N)$ contractions. 
The correct $U(1)$ limit follows from the vanishing of all
the covariant derivatives
$\nabla\ldots\nabla M=\nabla\ldots\nabla\eta=0$. The
crucial property to check is base-point independence. Note that, since
the expansion of $M_{AB}$ 
is $SO(dN^2)$ covariant, it is manifest that the action is
invariant under any matrix valued diffeomorphism. 
However, the bi-tensor $M_{AB}$ should also be a functional of the
$d$-dimensional 
metric, and its derivatives. This functional will be consistent with 
base-point independence, 
if under a shift 
in base-point, it is parallel transported in the $d$-dimensional
sense. As before, this is guaranteed if
\begin{equation}
    \epsilon^k \delta^{ab} \nabla_{kab} = \epsilon^k \nabla_k,
\end{equation}
on the tensor $M_{AB}$. For $M_{AB}=\eta_{AB}$ this is obviously so,
so that the result is indeed base-point independent.
Note that this is a truly non-trivial constraint that does
not follow from the fact that the Lagrangian $L_2$ is a scalar
quantity under matrix valued diffeomorphisms. 

Focussing on $M_{AB}$ alone, these results can
be extended to arbitrary order in $X$; at all orders $\nabla \ldots
\nabla M$ can be set to vanish consistent with the matrix-geometry
constraints. In particular, the non-trivial identity 
\begin{equation}
    \left .  \nabla_{[C}\nabla_{D]}M_{AB} \right |_{\bar X} =
    \mathrm{R}_{CDA}^{\phantom{CDA}Q} \left . M_{QB} \right |_{\bar X}
    + \mathrm{R}_{CDB}^{\phantom{CDB}Q}\left . M_{QA} \right |_{\bar
    X} = \mathrm{R}_{CDBA} + \mathrm{R}_{CDAB} = 0.
\end{equation}
and higher order analogues are satisfied. For these,
it is crucial that $M_{AB}$ evaluated at the base-point is equal
to the tangent space 
metric. In section \ref{sec:checks:-fourth-order}, where we
discuss consistency of the base-point independence approach for higher
derivative terms, we will see that these non-trivial identities do
impose constraints. 
We thus recover the two-derivative action (\ref{eq:secondordermg}). 
The lesson is that
base-point independence and the other constraints of matrix geometry
are automatically satisfied when we can put 
$\nabla\ldots\nabla M$ to zero. The power of the above argument by
generalization is that
the expansion of the action in RNC allows us to, almost, read off whether the
action is a candidate for the non-Abelian generalization of the
DBI-action.

\bigskip 
We now proceed with a number of checks and extensions of the action
for non-abelian D0-branes.  In section
\ref{sec:checks:-fourth-order} we show how the method just discussed
extends to actions fourth order in derivatives. But before that, we
perform two checks on our method. In the next section we prove that
the base-point independent potential term, conjectured in
\cite{DeBoer:2001uk}, turns into the kinetic term of
eq. (\ref{eq:secondordermg}) after T-duality. 
And in section
\ref{sec:checks:-non-transv}, we show how the 
matrix-geometry approach has a straightforward extension to spacetimes
with curvature along the brane (recall that the above approach
is only valid for transverse curvature).

\section{Checks: T-duality}
\label{sec:checks:-t-duality}
\setcounter{equation}{0}

One of the consequences of the non-abelian nature of the coordinates
$X$ is that new terms can be present in the action, proportional to
commutators, which have no $U(1)$ equivalent. Indeed for non-abelian
D0-branes in flat space, string theory tells us that at lowest order
in derivatives there is a potential equal to
\begin{eqnarray}
  \label{eq:18}
  V = -\frac{T\lam^2}{2}{\rm Tr}([X^i,X^j][X_i,X^j])~.
\end{eqnarray}
The form of the potential is dictated by consistency with 
T-duality. Under this stringy symmetry the potential and kinetic term
are exchanged. T-duality holds for any spacetime with isometries, and
the curved space analogues of both the potential and the kinetic term
should be consistent with the duality. In addition to constructing a
base-point independent kinetic term, in \cite{DeBoer:2001uk} we also
put forward a potential for D0-branes in a curved
background. This conjectured potential passed a strong consistency
test. It satisfied the
non-trivial D-geometry constraint that fluctuations around a
diagonal background have masses proportional to geodesic lengths. We
will now show that the conjectured form of the potential
reproduces the kinetic term after a T-duality transformation. This is
strong confirmation that our guess is correct.

To generalize the expression \eqref{eq:18} 
to curved space, we need an analogue of
the vector $\dot{X}^I$, which can be contracted with the vielbein
$E_I^A$. Define the ``commutation'' operator
\begin{eqnarray}
  \label{eq:19}
  \cD(X)^{icb;ad} \equiv \delta^{ab} X^{icd}-\delta^{cd}X^{iab}~.
\end{eqnarray}
Acting on a matrix $M_{da}$ it returns the commutator
\begin{eqnarray}
  \label{eq:20}
  \cD(X)^{icb;ad}M_{da} = [X^i,M]^{cb}~.
\end{eqnarray}
Commutators obey the Leibniz rule and act as a derivation on the space
of matrices. Analogous 
to the standard time derivative $\dot{X}$, we expect any $X^{iab}$
appearing inside a commutator to transform as a vector under
matrix coordinate (i.e. base-point) transformations. 
The matrix-valued ``commutation operator'' $\cD(X)^{I,ad}$ can
therefore be pushed forward to the tangent space with the $SO(dN^2)$
vielbein. Supporting and {\em consistent with} the notion 
that the ``commutation''
operator is the covariant building block of the potential, is the
expression for 
flat space potential \eqref{eq:18} in terms of $\cD(X)^{I,ab}$. A
small calculation shows that it is equivalent to four
building blocks
contracted with exactly twice the $SO(dN^2)$ metric
\cite{DeBoer:2001uk}. 
\begin{eqnarray}
  \label{eq:23}
  V_{flat} &=& -\frac{T\lam^2}{4}
  \eta_{IK}\eta_{JL}
\cD(X)^{I,ab}\cD(X)^{J,bc}\cD(X)^{K,cd}\cD(X)^{L,da} 
  \non
&=& -\frac{T\lam^2}{4} \eta_{IK}\eta_{JL} {\rm
  Tr}\cD(X)^{I}\cD(X)^{J}\cD(X)^{K}\cD(X)^{L}~. 
\end{eqnarray}
In the last line the trace is only over the explicit $U(N)$ indices of
the matrix valued $SO(dN^2)$ vector $\cD(X)^{I,ab}$.

The generalization to curved space is now straightforward. We simply
insert the appropriate number of 
vielbeins into the flat space potential:
\begin{equation}
\begin{split}
\label{eq:24}
    V_{curved} &= -\frac{T\lam^2}{4} \eta_{AC}\eta_{BD}
    E^A_IE^B_JE^C_KE^D_L{\rm
      Tr}\cD(X)^{I}\cD(X)^{J}\cD(X)^{K}\cD(X)^{L} 
\\ &= -\frac{T\lambda^2}{4} (\eta_{AC}\eta_{BD}
  + \frac{4}{12} \mathrm{R}_{C(PQ)A}X^PX^Q\eta_{BD}) {\rm
    Tr}\cD(X)^{A}\cD(X)^{B}\cD(X)^{C}\cD(X)^{D} + \mathcal{O}(\nabla^2)\\ &=
    -\frac{T\lambda^2}{4}
    (\tr([X^i,X^k][X^j,X^l])\delta_{ij}\delta_{kl}+\frac{1}{3}
    \mathrm{R}_{i(kl)j} \delta_{mn} \str(X^kX^l[X^i,X^m][X^j,X^n])) +
    \ldots,
\end{split}
\end{equation}
The last two steps shows that 
in the linearized approximation 
it correctly reproduces the
symmetrized result from
\cite{taylorvanraamsdonk}, as is expected. 

Under T-duality the ``parallel'' part of the 
curved space potential \eqref{eq:24} must
transform into the kinetic term \eqref{eq:secondordermg}. To check
this, assume that the $d$-dimensional geometry is a product of a
$(d-1)$-dimensional piece times a circle along the direction $i=9$. 
This is not sufficient to test the full nature of
T-duality, but in this situation everything is tractable and
certainly should work. We thus have the following expression 
for the $dN^2$ metric:
\begin{eqnarray}
  \label{eqa:3}
  G_{i\alp\bet,j\gam\del}(X^i) = \left(\mtrix{
  G_{\mu\alp\bet,\nu\gam\del}(X^{\rho}) & 0 \cr 0 &
  \del_{\alp\del}\del_{\bet\gam}} \right)~~~~~\mu,\nu,\rho =
  1,\ldots,d-1.
\end{eqnarray}
The expression in the lower right corner is a simply a consequence of
the non-trivial form of the flat-space metric:
\begin{eqnarray}
  \label{eqa:4}
   G^{flat}_{i\alp\bet,j\gam\del} = \eta_{i\alpha\beta,j\gamma\delta}
   = \delta_{ij}\delta_{\alpha\delta}\delta_{\beta\gamma}~.
\end{eqnarray}
Because we have chosen a direct product form for the space time, the
tangent space decomposes trivially:
\begin{eqnarray}
  \label{eq:26}
  \eta_{a\alp\bet,b\gam\del}^{d-dim} =
                      \left(\mtrix{\eta_{m\alp\bet,n\gam
                      \del }^{(d-1)-dim} &0 \cr
                      0 &
                      \delta_{\alp\del}\delta_{\beta\gam}}\right)~. 
\end{eqnarray}
In particular, the component of the 
vielbein $E^{A}_I$ along the circle is 
\begin{eqnarray}
  \label{eq:25}
  E_{9\alp\bet}^{a;\eps\zet} =
  \delta^a_9\delta^{\eps}_{\alp}\delta^{\zet}_{\bet} ~.
\end{eqnarray}

In this background the potential splits into three parts. One has
no tangent-vectors lying along the isometry direction; this will
become the potential in the T-dual case. A  second term has {\em all}
components along the circle: since we know the flat limit corresponds
to the commutator squared, this term will vanish. The crossterm with
half the components along the circle is the interesting part. Using
that the vielbein is trivial in the ninth direction, it equals
\begin{eqnarray}
  \label{eq:28}
  V_{curv}^{cross} = -\frac{T\lam^2}{2} \eta_{9ab,9cd}\eta_{BD}
  E^B_JE^D_L{\rm Tr}\cD(X)^{9ab}\cD(X)^{J}\cD(X)^{9cd}\cD(X)^{L}~. 
\end{eqnarray}
The triviality of the vielbein allows the use of the following contraction
identity, 
\begin{eqnarray}
  \label{eq:27}
  \eta_{IJ}\cD(X)^{I,ad}\cD^{J,ef} = 2 (X^k)^{af}(X_k)^{ed}
  -\delta^{af} (X^2)^{ed} -\delta^{ed}(X^2)^{af},
\end{eqnarray}
in the direction of the circle. We get
$$
   V_{curv}^{cross} = -{T\lam^2} \eta_{BD}E^B_JE^D_L
   \left[{\rm Tr}\left(X^9\cD(X)^{J}\right){\rm
   Tr}\left(X^9\cD(X)^{L}\right) - {\rm Tr}\left(\cD(X)^{J}\right){\rm
   Tr}\left(X^9X^9\cD(X)^{L}\right) \right]~. 
$$
Due to 
the defining property \eqref{eq:20} of the commutation operator, 
the last term vanishes as Tr$(D(X)^I)=[X^i,\one] =
0$. The remaining term yields
\begin{eqnarray}
  \label{eq:30}
V_{curv}^{cross} = -{T\lam^2}
\eta_{BD}E^B_{j\alp\bet}E^D_{\ell\gam\del}
\left([X^j,X^9]^{\alp\bet}[X^{\ell},X^9]^{\gam\del}\right) 
~. 
\end{eqnarray}
Upon using the standard T-duality rule which replaces commutators with
derivatives, 
$$
i\lam[X^9,\cF(X)] \rar \pa_9\cF~,
$$ 
we recover exactly the
kinetic term \eqref{eq:secondordermg}. Notice
that the vielbeins have basically just gone along for the ride. The
proof of T-duality in a flat background would be identical. This shows
the power of the matrix-geometry approach.

\section{Checks: Non-transverse metrics}
\label{sec:checks:-non-transv}
\setcounter{equation}{0}

A second crucial test which the base-point independent action for
non-abelian D0-branes must pass, is that it must be able to account
for gravitational polarization along the worldvolume in addition to
purely transverse curvature. So far we have only dealt with the latter
situation, both for simplicity as well as the string
theory indication 
that only the transverse coordinates get promoted
to $U(N)$-valued matrices. This is not a very satisfactory situation. 
Most interesting metrics, e.g. Schwarzschild, (A)dS, have curvature
in the timelike tangential direction. Moreover, from a
diffeomorphic perspective it is very strange. It appears to conflict 
with the idea of general coordinate invariance, since some
directions are said to be more special than others, based on data not
intrinsic to the space-time. Introducing an extension to a physical
gauge choice, we will see that this conflict is spurious. 
This important conclusion has been instrumental in providing
evidence that there is a gravitational analogue to the Myers effect
\cite{deBoer:2002we}.\footnote{The results in this 
section were obtained
  together with Eric Gimon.} 
  
There is one obvious answer to deal with more general metrics, that
ensures an action that is fully diffeomorphism invariant. That answer
is to start with the $dN^2+p+1$ dimensional NL$\sigma$M,
\begin{equation}
\label{eqb:1}
 S = \int d\xi^{p+1} \sqrt{\det(G_{MN} \frac{\partial X^M}{\partial
\xi^a} \frac{\partial X^N}{\partial \xi^a})}~,
\end{equation}
where $M$ is now the multi-index ${\hat{M}=m\alpha\beta,a}$; demand
that $G_{MN}(X)$ is a functional of the metric $g_{mn}$ and its
derivatives, and solves the set of constraints
\begin{itemize}
\item[(a)] The $dN^2+p+1$ dim Riemann tensor $R_{IJKL}$, its covariant
  derivatives, etc. have all the
usual properties.
\item[(b)] The action is a single trace over the $U(N)$ indices.
\item[(c)] The action has the correct linearized form and $U(1)$ limit.
\item[(d)] The action is base-point independent: i.e. 
$\delta^{\alpha\beta}\nabla_{m\alpha\beta} = \nabla_m$.
\end{itemize}
Of course to set up the system of constraints algebraically one needs
to be careful whether the index $M$ is orthogonal to the brane (in the
$\hat{M}$ direction) or parallel (in the $a$ direction).

This procedure emphasizes the dichotomy which conflicts with
diffeomorphism invariance that 
directions perpendicular to the
brane are treated differently than those parallel.  
For a single D-brane, we know this
is a fake problem. Introducing a wordline metric, we can "undo" the
physical gauge choice $\tau=x^0$; the extra degree of freedom $X^0(\tau)$ 
is compensated by the additional worldsheet diffeomorhism
symmetry. Up to two derivatives, the action is
then exactly
the same as before, except that the range of indices now also includes
the tangential directions. 

For non-abelian D0-branes the conflict due to this dichotomy 
is far more accute: 
from its origins in string theory, 
we expect the tangential
coordinates to be commutative, while diffeomorphism invariance tells
us that they should be of the same non-commutative nature as the
transverse coordinates. Physically, however, we expect that a
``physical gauge'' solution should
also exist for $N$ D-branes. Consider e.g. the example
of $N$ static D1-branes wrapping the equator of a 2-sphere. Geometrically the
normal and tangential directions are equivalent; yet the natural
construction advocated above seems to say the opposite. This cannot be
true. Simply rotating the system around a fixed point on the
equator, should leave everything invariant. This argues that "undoing"
the physical gauge choice should also give tangential matrix valued
coordinates. Thus we have a straightforward 
guess for the solution to our dilemma. Up to two derivatives the
action is simply the same as before (\ref{eq:secondordermg}) but with
the range of indices including the tangential directions. And the system has
enough symmetry that we may choose a physical gauge
\begin{equation}
\label{eqb:2}
 X^{\parallel \alpha \beta}= \xi^{\parallel} \delta^{\alpha\beta}~.
\end{equation}
It seems difficult to justify this procedure from a world-sheet
point of view. Nevertheless geometrically it 
seems to be the 
most natural thing to do.
We simply propose that this is an inherent
part of matrix-diffeomorphisms. 
Support for this proposal is that the final action will obey all
the 
correct symmetries. 

Of course, so would the action constructed by the
dichotomous approach proposed below eq. (\ref{eqb:1}). 
This is as it should be. It is
not so hard to show, that the actions resulting
from either imposing a matrix-physical gauge or the "natural" p-brane
procedure above are the same. This is the evidence in support of the fact that
in matrix-geometry enough symmetries exist, to fix the matrix-physical
gauge. To show equivalence, we use the single trace property. Let us first
look at the Riemann tensor. We use tilded indices to denote the
$(d+p+1)N^2$ directions, greek indices in the middle of the alphabet 
for the underlying $d+p+1$
dimensional spacetime, greek indices in the beginning of the alphabet
for $U(N)$ indices, $a,b$'s and $m,n$'s for the tangential and
normal directions respectively, capital indices for the $dN^2+p+1$
NL$\sigma$M, splitting in $dN^2$ $\hat{M}$'s $= m\alpha\beta$ and
$p+1$ $a$'s.

We start from the $(d+p+1)N^2$ solution to the constraints for the
Riemann tensor:
\begin{equation}
\label{eqb:3}
 R_{\tilde{M}\tilde{N}\tilde{P}\tilde{S}} =
 R_{\mu\alp_1\alp_2,\nu\beta_1\beta_2,\rho\gam_1\gam_2,\sig\delta_1\del_2}
 = \sum_{n=1}^6
T^{(n)}_{\mu\nu\rho\sigma} \Delta_{p_n(\alpha\beta\gamma\delta)}~,
\end{equation}
where $p_n(\alpha\beta\gamma\delta)$ is the $n$th permutation and
\begin{equation}
\label{eqb:4}
\Delta_{\alpha\beta\gamma\delta} \equiv
\delta_{\alpha_2\beta_1}\delta_{\beta_2\gamma_1}\delta_{\gamma_1\delta_2}
\delta_{\delta_1 \alpha_2}
\end{equation}
is the cyclic contraction. Hence the total number of inequivalent
permutations $p_n(\alpha\beta\gamma\delta)$ is 6 for a 4-tensor; or
$(n-1)!$ for $n$ indices. $T^{(n)}_{\mu\nu\rho\sig}$ is functional of
the $d+p+1$-dimensional Riemann tensor. Now we impose the physical gauge by
contracting those indices in $\Delta_{\alpha\beta\gamma\delta}$ with
$\delta^{\alpha_1\alpha_2}$ that correspond with tangential
directions. The answer is obvious, the contraction of
$\Delta_{\alpha\beta\gamma\delta}$ with $\delta^{\alpha_1\alpha_2}$
yields a $\Delta$ tensor with one less index. Or specifically
\begin{eqnarray}
\label{eqb:5}
 R_{\tilde{A}\tilde{N}\tilde{P}\tilde{S}}|_{phys.gaug} = \sum_{n=1}^2
T^{(n)}_{anrs} \Delta_{p_n(\beta\gamma\delta)}~, \non
R_{\tilde{A}\tilde{B}\tilde{P}\tilde{S}}|_{phys.gaug} = \sum_{n=1}^1
T^{(n)}_{abrs} \Delta_{p_n(\gamma\delta)} ~,\non
R_{\tilde{A}\tilde{N}\tilde{B}\tilde{S}}|_{phys.gaug} = \sum_{n=1}^1
T^{(n)}_{anbs} \Delta_{p_n(\beta\delta)} ~,\non
R_{\tilde{A}\tilde{B}\tilde{C}\tilde{S}}|_{phys.gaug} = T_{abc\sigma}~.
\end{eqnarray}
These are exactly the first set of constraints one would write down
for the $dN^2+p+1$ dimensional dichotomous sigma model. And 
there is a
beautiful corollary to this Riemann tensor equivalence of the two approaches:
the base point independence constraint ({\em in other words
diffeomorphism invariance}), $\delta^{\alpha\beta}
\nabla_{\mu\alpha\beta} = \nabla_{\mu}$, guarantees that this also
holds for all derivatives of the Riemann tensor. Thus the physical
expectation that diffeomorphism invariance should treat all
coordinates equally is born out, once we "undo" the physical gauge
choice $X^{\parallel} = \xi^{\parallel}\delta^{ab}$.

\section{Balances: Higher order corrections: Fourth order in $\dot X$}
\label{sec:checks:-fourth-order}
\setcounter{equation}{0}

We have thus seen that the base-point independent action including the
potential is consistent with T-duality and that the method itself
captures diffeomorphism invariance fully in that it is
extensible to 
non-transverse curvature for actions up to two derivatives. We
conclude here with an application of the base-point independence
approach to the next order in derivatives. This will illustrate the
universality of our method and we will
show at this order the symmetrized trace prescription is no longer 
consistent with
base-point independence, even if we treat gravity at the linearized level.
We conclude that the symmetrized trace
approximation is that of linearized gravity {\em up to two
  derivatives.}

In section \ref{sec:applying-base-point} we saw
that the covariant expansion is a straightforward approach to 
determine the crucial
properties of the action. We will therefore 
pursue this route also for the action $L_4$, the part of the
DBI-action of fourth order in $\dot X$. Before doing that, let us show
explicitly why the symmetrized trace approximation starts to fail at
this order. If the
action would be the symmetrized
trace, then it could be written as:
\begin{equation}
\begin{split}
    L_4 &= -\tfrac{1}{8} \str(g_{ij}(X)g_{kl}(X){\dot X}^i{\dot
    X}^j{\dot X}^k{\dot X}^l) \\ &= -\tfrac{1}{8} (
    \delta_{ij}\delta_{kl}\str({\dot X}^i{\dot X}^j{\dot X}^k{\dot
    X}^l) + \tfrac{2}{3}\mathrm{R}_{imnj}\delta_{kl}\str(X^mX^n{\dot
    X}^i{\dot X}^j{\dot X}^k{\dot X}^l))+\ldots.
\end{split}
\end{equation} 
Substituting the base-point transformation (a constant shift
$\epsilon$) (the explicit expression follows from the
geodesic equation in matrix space, which follows from
$L_2$.) :
\begin{equation}
    \Delta X_i = \epsilon_i +
    \tfrac{1}{6}\epsilon^k\sym(Z^{p_1}Z^{p_2})\mathrm{R}_{p_1(ki)p_2},
\end{equation}
we get for the variation (schematically and up to first order in the
Riemann tensor):
\begin{equation}
    \Delta L_4 \propto \epsilon \mathrm{R}\str({\dot Z}^3\sym({\dot Z}
    Z)) + \epsilon \mathrm{R}\str({\dot Z}^4 Z).
\end{equation}
From this we see that the two combinatorial structures cannot be
combined, hence the symmetrized trace prescription does not yield a
base-point independent action.

To determine what is the correct ordering, 
we follow the same steps as before. Our
starting point is an action of order four in one-forms $\Pi^A$ contracted
with an arbitrary symmetric four-tensor $M_{ABCD}(X)$:
\begin{equation}
    L_4 = -\tfrac{1}{8} M_{ABCD}(X)\Pi^A\Pi^B\Pi^C\Pi^D~.
\end{equation}
Expanding in RNC up to second order in $X$, we find
\begin{equation}\begin{split}
    -8 L_4 &= \left \{ \left . M_{ABCD} \right |_{\bar X} + \left
     . \nabla_FM_{ABCD} \right |_{\bar X}X^F \right. \\ &+ \left
     . \tfrac{1}{2} \left . (\tfrac{4}{3} M_{QBCD}
     \mathrm{R}^Q_{\phantom{Q}EFA} + \nabla_E\nabla_FM_{ABCD}) \right
     |_{\bar X}X^EX^F \right \} {\dot X}^A{\dot X}^B{\dot X}^C{\dot
     X}^D.
\end{split}\end{equation}

The lesson from section \ref{sec:applying-base-point} is whether 
we can put $\nabla \ldots \nabla \left . M \right |_{\bar X} = 0$? 
This would ensure the correct $U(1)$ limit and base-point independence.
Suppose we could. In that case the action would be:
\begin{equation}
    -8 L_4 = \left \{\left . M_{ABCD} \right |_{\bar X} +\left
     . \tfrac{2}{3} M_{QBCD} \mathrm{R}^Q_{\phantom{Q}EFA} \right
     |_{\bar X} X^EX^F \right \} {\dot X}^A{\dot X}^B{\dot X}^C{\dot
     X}^D\
\end{equation}
From the flat space limit,\footnote{The flat space limit follows
  indirectly from explicit string computations which show that the
  non-abelian DBI-action at order $F^4$ is given by the
  symmetrized trace \cite{Bergshoeff:1986jm}. This breaks down at
  higher orders; see \cite{Sevrin:2003vs} for the latest status.}
\begin{equation}
    -8 L_4^{flat} = \left . M_{ABCD} \right|_{\bar X} {\dot X}^A{\dot
     X}^B{\dot X}^C{\dot X}^D = \str({\dot X}^i{\dot X}^j{\dot
     X}^k{\dot X}^l)\delta_{ij}\delta_{kl},
\end{equation}
 we learn that:
\begin{equation}
     \left. 
M_{ABCD} \right|_{\bar X} \equiv \eta_{ABCD} =
    \delta_{ab}\delta_{cd} \Sig_{\alp_1\bet_1\ldots\alp_4\bet_4} \neq
    \eta_{AB}\eta_{CD},
\end{equation}
Note that the single trace requirement  means
that the tensor $M$ is not proportional to two tangent metrics. This
will be important.

Are all the requirements (a)-(d) of matrix geometry satisfied? 
Since we've assumed all covariant derivatives on $M$ can be put to
zero we have obtained the right $\mathrm{U}(1)$ limit. The 
single trace condition is met, by construction. What about the focus
of this article: base-point independence?
This is also guaranteed
if we can truly 
put all covariant derivatives on $M$ to zero (evaluated in
the base-point). It therefore remains to test this assumption. 
Because $M$ is
not the tensor product of the metric tensor, it is not 
clear that a covariantly constant $M_{ABCD}$ exists. The litmus tests
are the identities involving commutators of covariant derivatives on
$M_{ABCD}$; the simplest being
\begin{equation}
\begin{split}
    \nabla_{[E}\nabla_{F]} \left . M_{ABCD} \right |_{\bar X} &= 2
(\mathrm{R}_{EFA}^{\phantom{EFA}Q} M_{QBCD} +
\mathrm{R}_{EFB}^{\phantom{EFB}Q} M_{QACD} \\ &\left . +
\mathrm{R}_{EFC}^{\phantom{EFC}Q} M_{QABD}+
\mathrm{R}_{EFD}^{\phantom{EFD}Q} M_{QABC} ) \right |_{\bar X}.
    \label{eq:covdiffid}
\end{split}
\end{equation}
This directly 
shows that a covariantly constant $M$, not proportional to the metric,
is not consistent since the right-hand side of the equation 
will not evaluate to zero. Recall that in the two derivative case, we
explicitly tested this. There $M_{AB}$ equalled the metric, and 
we did not have this
problem since
\begin{equation}
\label{eq:31}
    \left .  \nabla_{[C}\nabla_{D]}M_{AB} \right |_{\bar X} =
    \mathrm{R}_{CDA}^{\phantom{CDA}Q} \left . M_{QB} \right |_{\bar X}
    + \mathrm{R}_{CDB}^{\phantom{CDB}Q}\left . M_{QA} \right |_{\bar
    X} = \mathrm{R}_{CDBA} + \mathrm{R}_{CDAB} = 0.
\end{equation}
The mathematical cause for the inconsistency of a covariantly constant
$M_{ABCD}$ is the single trace requirement. It was responsible for the
fact that $M_{ABCD}$ could not
equal two tangent space metrics.\footnote{
One might object that the identity \eqref{eq:31}
appears to be irrelevant as in the
action $\nabla_E\nabla_FM_{ABCD}$ is contracted with the symmetric combination
$X^EX^F$. The remaining discussion will show why this is not so.}

Lacking a deeper insight in the requirements on $M_{ABCD}$, we are
forced to check 
base-point independence by hand. 
Recall that the base-point transformation is given by:
\begin{equation}
    \Delta X_i = \epsilon_i +
    \tfrac{1}{6}\epsilon^k\sym(Z^{p_1}Z^{p_2})\mathrm{R}_{p_1(ki)p_2}.
\end{equation}
Using this transformation we calculate the variation of
$L_4$ with $M_{ABCD}\equiv \eta_{ABCD}$:
\begin{eqnarray}
    -8 L_4 &=&
    \eta_{ABCD} E^A_I{\dot X}^IE^B_J{\dot X}^JE^C_K{\dot
    X}^KE^D_L{\dot X}^L \non
&=&  \delta_{ij}\delta_{kl} \str({\dot X}^i{\dot X}^j{\dot
    X}^k{\dot X}^l) + \tfrac{1}{3} \mathrm{R}_{i(kl)j}\delta_{mn}
    \str({\dot X}^i{\dot X}^m{\dot X}^n\sym({\dot X}^jX^kX^l))~, \\
    -\Delta L_4 &=& \tfrac{1}{8} \tfrac{4\cdot2}{3\cdot 2} \epsilon^k
    \mathrm{R}_{\beta(k\alpha)p_2} \str({\dot Z}^\alpha {\dot Z}^2
    \sym({\dot Z}^\beta Z^{p_2})) + \tfrac{2}{3\cdot
    8}\mathrm{R}_{k(\alpha\beta)p_2}\epsilon^k\str({\dot
    Z}^\alpha{\dot Z}^2\sym({\dot Z}^\beta Z^{p_2})) \non 
    & =&
    \tfrac{1}{6} \epsilon^k \str({\dot Z}^\alpha {\dot Z}^2 \sym({\dot
    Z}^\beta Z^{p_2})) \left \{ \mathrm{R}_{\beta(k \alpha)p_2} +
    \tfrac{1}{2} \mathrm{R}_{k (\alpha\beta)p_2} \right \}.
\end{eqnarray}
Note that $\sym(\ldots)$ expressions are treated as one block within
the symmetrized trace.
Using (note $A_{(ab)} = A_{ab}+A_{ba}$),
\begin{displaymath}
    \mathrm{R}_{\alpha(p_1\beta)p_2} = -\tfrac{1}{2}
    \mathrm{R}_{p_1(\alpha\beta)p_2}+\tfrac{3}{2}
    \mathrm{R}_{\alpha\beta p_1p_2},
\end{displaymath}
this simplifies to (with the notation $A_{p_1\ldots p_n} =
A_{1\ldots n}$):
\begin{equation}
    \Delta L_4 = -\tfrac{1}{4}
    \epsilon^6\mathrm{R}_{3456}\delta_{12}\str({\dot Z}^1{\dot
    Z}^2{\dot Z}^3\sym({\dot Z}^4Z^5)) \neq 0.
\end{equation}
So the proposed action is indeed not base-point
independent. To see that the single trace requirement is responsible
--- and hence the 
correlated inconsistency of choosing $M_{ABCD}(X)=\eta_{ABCD}$ --- 
let us look at the corresponding calculation for $L_2$. The steps are
analogous and the $L_2$ result comes down to removing the
$\delta_{12}{\dot Z}^1{\dot Z}^2$ part in the $L_4$ result 
and adjusting some factors:
\begin{equation}
    \Delta L_2 \propto \epsilon^6 \mathrm{R}_{3456} \str({\dot
    Z}^3\sym({\dot Z}^4Z^5)).
\end{equation}
Here we can make use of the identity:
\begin{equation}
    \str(ABCD\ldots) = \tr(A\sym(BCD\ldots)).
    \label{eq:strid}
\end{equation}
This allows us to write $\Delta L_2$ as:
\begin{equation}
    \Delta L_2 = \epsilon^6\mathrm{R}_{3456}\str({\dot Z}^3{\dot
    Z}^4Z^5),
\end{equation}
which is obviously zero. In the case of $\Delta L_4$ the identity in
equation \ref{eq:strid} is of no use to us, because of the extra
$\delta_{12}{\dot Z}^1{\dot Z}^2$ factor. Had we not insisted on a
single trace result and used $M_{ABCD}=\eta_{AB}\eta_{CD}$, 
then the variation of
$\Delta L_4$ would have had the structure:
\begin{equation}
    \Delta L_4 = \Delta L_2 \delta_{12}\str({\dot Z}^1{\dot Z}^2) = 0.
\end{equation}
So, as claimed, the single trace property spoils base-point
independence. As a result of this we confirm the importance of the identity
\eqref{eq:31}. $\nabla_E\nabla_F 
\left . M_{ABCD} \right |_{\bar X}$ should not be zero if we insist on
base-point independence.

Fortunately is it not terribly difficult to find an correction term to
$L_4$ that renders the action base-point independent while keeping the
correct $\mathrm{U(1)}$ limit. 
One possible answer is:
\begin{equation}
    L^C_4 = \alpha \mathrm{R}_{1356}\delta_{24}\str({\dot X}^1{\dot
    X}^2\sym({\dot X}^4X^5)\sym({\dot X}^3X^6)).
\end{equation}
Since $R_{1356}$ is antisymmetric in
1 and 3, this result vanishes in the $U(1)$ limit.
The variation of $L^C_4$ equals:
\begin{equation}
\begin{split}
    \Delta L^C_4 &= \alpha \mathrm{R}_{1356}\delta_{24}\str({\dot
Z}^1{\dot Z}^2{\dot Z}^4\epsilon^5\sym({\dot Z}^3Z^6)) \\ &+ \alpha
\mathrm{R}_{1356}\delta_{24}\str({\dot Z}^1{\dot Z}^2\sym({\dot
Z}^4Z^5){\dot Z}^3\epsilon^6) \\ &= \alpha \epsilon^6
\mathrm{R}_{1465}\delta_{24}\str({\dot Z}^1{\dot Z}^2{\dot
Z}^3\sym({\dot Z}^4Z^5)) \\ &= 4\alpha \Delta L_4.
\end{split}
\end{equation}
Requiring 
base-point independence determines the constant $\alpha$ to be
$-\tfrac{1}{4}$.

The obvious next question is what is the value for 
$\nabla\ldots\nabla \left . M_{ABCD}
\right |_{\bar X}$ that corresponds to $L^C_4$. 
Since there is no $\mathcal{O}(X)$
term in the $\mathrm{U}(1)$ limit, we will try to maintain this
property for the non-abelian case: we require
that there is no $\nabla_F \left . M_{ABCD} \right |_{\bar X} X^F$
term in the action. For the $\nabla_E\nabla_F \left . M_{ABCD} \right
|_{\bar X} X^EX^F$ part we can only determine the part symmetric in
$EF$ and $ABCD$ from $L^C_4$. It is
determined by:
\begin{equation}\begin{split}
    &\nabla_E\nabla_F \left . M_{ABCD} \right |_{\bar X} X^EX^F{\dot
    X}^A{\dot X}^B{\dot X}^C{\dot X}^D = 4
    \mathrm{R}_{1456}\delta_{23} \str({\dot X}^1{\dot X}^2\sym({\dot
    X}^4X^5)\sym({\dot X}^3X^6)) 
\end{split}\end{equation}
We write this schematically as (see appendix \ref{app:notation} for the
notation):
\begin{equation}
    \nabla_E\nabla_F \left . M_{ABCD} \right |_{\bar X} X^EX^F{\dot
    X}^A{\dot X}^B{\dot X}^C{\dot X}^D = \tfrac{1}{48}
    s_{(ABCD)(EF)}X^EX^F{\dot 
    X}^A{\dot X}^B{\dot X}^C{\dot X}^D
\end{equation}
with  
\begin{equation}
s_{ABCDEF}
    = s_{123456} = 4 \mathrm{R}_{1456}\delta_{23}
    \Sigma_{1278}S^7_{45} S^8_{36},
    \label{eq:explicitS}
\end{equation}
We already showed explicitly that the action thus obtained is
base-point independent. However, consistency of the more
general approach demands that 
base-point independence can also be shown by proving the
following:
\begin{equation}
    \epsilon^i \delta^{ab} (\nabla_{iab}\nabla_F\left. M_{ABCD} \right
    |_{\bar X}) = \epsilon^i (\nabla_i \nabla_F \left . M_{ABCD}
    \right |_{\bar X}).
    \label{eq:bpiM}
\end{equation}
In order to check this we have to know what the right-hand
part of the previous equation is. It corresponds to the
variation of the $\nabla_F \left . M_{ABCD} \right |_{\bar X} X^F$
term. Since this term vanishes, we conclude that
the right-hand side of equation \ref{eq:bpiM} should be zero 
in the
action. Recall that 
$\nabla_E\nabla_F \left
. M_{ABCD} \right |_{\bar X}$ can receive several contributions, of
which only the $S$ part contributes to the action. Of course, 
if it vanishes as a
single formal tensor, that would be the more satisfactory.  Using the
explicit representation for $s_{ABCDEF}$ (equation 
\ref{eq:explicitS}), we find for the left-hand side of eq. (\ref{eq:bpiM}):
\begin{equation}
\begin{split}
    &\epsilon^i \delta^{ab} S_{1234\,iab\,6} = \epsilon^i \delta^{ab}
    (\mathrm{R}_{14i6} \delta_{23} \Sigma_{1278} S^7_{4\,ab}
    S^8_{36}+\ldots) \\ &= \epsilon^i ( \mathrm{R}_{14i6} \delta_{23}
    \Sigma_{1248} S^8_{36} + \ldots) \\ &\propto \epsilon^i
    (\mathrm{R}_{14i6} \delta_{23} \Sigma_{1248} S^8_{36} +\epsilon^i
    \mathrm{R}_{41i6} \delta_{23} \Sigma_{1248} S^8_{36} + \ldots) =
    0.
\end{split}
\end{equation}
This confirms the base-point independence. The other contributions to
$\nabla\nabla M$, such as the parts anti-symmetric in $E,F$, are of no
concern since these vanish in the action. It is, however, 
possible to add other corrections terms to $M$ such that
equation \ref{eq:bpiM} is zero, even as a formal tensor.

Presumably hese results can be generalized to arbitrary order in the velocity
$\dot X$. However, the extension to
higher order in $X$ is far from trivial due to the complications from
the inability to put $\nabla
\ldots \nabla M$ to zero in matrix geometry. 
For every order in $X$ we have to find the
appropriate tensor $\nabla \nabla \ldots \nabla \left . M \right
|_{\bar X}$, which is beyond our present capability. The fact that we
are able to do so for fourth order in derivatives does give confidence
that this is possible. Our computation does 
explicitly show that
at the first corrective order in derivatives the symmetrized trace
approximation is no longer consistent with the single trace
requirement in the presence of gravity.


\bigskip

\noi 
{\bf Acknowledgments:} We sincerely wish to thank Eric Gimon, who 
collaborated with us on part of this work. We also 
thank both the hosts and
participants of the Amsterdam Summer Workshop  
as well as the Aspen
Center for Physics for the hospitality and enlightening
discussions. 
JdB and JW thank Stichting FOM for support. 
KS gratefully acknowledges support from DOE grant
DE-FG-02-92ER40699.

\appendix
\section{Notation}
\label{app:notation}
\setcounter{equation}{0}

\noindent \hspace{-.1in}-\hspace{.05in}\emph{Riemann Normal
  Coordinates} 

\noindent We write fluctuations in a metric $g_{ij}$ as $h_{ij}$:
$g_{ij} = \eta_{ij} + h_{ij}$. The moments of $g$ evaluated at the
base-point $p$: 
\begin{displaymath}
    \partial_{k_1}\ldots\partial_{k_n} h_{ij} = \frac{(n-1)}{(n+1)!}
    \nabla_{(k_1\ldots}\nabla_{k_{n-2})} \mathrm{R}_{k_{n-1}|ij|k_n} +
    i\leftrightarrow j = \frac{(n-1)}{(n+1)} \mathrm{R}_{k_1ijk_2\ldots k_n} +
    i\leftrightarrow j.
\end{displaymath}
Note that parenthetical symmetrization of $n$ objects has weight $n!$
instead of the usual weight 1, i.e.:
\begin{displaymath}
    \nabla_{(k_1}\nabla_{k_2)} = \nabla_{k_1}\nabla_{k_2} +
    \nabla_{k_2}\nabla_{k_1}.
\end{displaymath}

\medskip
\noindent \hspace{-.1in}-\hspace{.05in}\emph{Matrix Geometry}

\noindent Capital letters refer to a multi-index notation in which a
matrix $X^i$ is represented as $X^I=X^{i\alpha\beta}$. Sometimes it is
easier to work in a local-Lorentz frame, in this case the matrix is
written as: $X^A=X^{\imath ab}$. The vielbein relating the $X^I$ and
the $X^A$, 
\begin{displaymath}
    X^A = E^A_I X^I \Rightarrow X^{\imath ab} = \sum_{i\alpha\beta}
    E^{\imath ab}_{i\alpha\beta} X^{i\alpha\beta},
\end{displaymath}
has a convenient flat space representation:
\begin{displaymath}
    \left . E^A_I \right |_\text{flat} = \delta^A_I = \delta^\imath_i
    \delta^a_\alpha \delta^b_\beta.
\end{displaymath}
The flat metric is defined in such a way that:
\begin{displaymath}
    \mathrm{tr}(X^iX^j)\delta_{ij} = \eta_{IJ} X^IX^J \Rightarrow
    \eta_{IJ} = \eta_{i\alpha\beta,j\gamma\delta} =
    \delta_{ij}\delta_{\alpha\delta}\delta_{\beta\gamma}.
\end{displaymath}
The metric for curved space is the given by:
\begin{displaymath}
    G_{IJ} = E^A_IE^B_J\eta_{AB}
\end{displaymath}
The Riemann tensor in matrix geometry evaluated at the base-point $X=p
$ is:
\begin{displaymath}
    R_{IKLJ}(p) X^AX^BY^KY^L = R_{iklj}(p)\mathrm{Str}(X^iX^jY^kY^l).
\end{displaymath}
Here $\mathrm{Str}$ is the symmetrized trace (symmetrization has
weight one).

\medskip

\noindent\hspace{-.1in}-\hspace{.05in}\emph{Miscellaneous}

\noindent The object $\Sigma$ is defined such that:
\begin{displaymath}
    \Sigma_{a_1b_1,\ldots,a_nb_n} O^{i_1a_1b_1}\cdots O^{i_na_nb_n} =
    \mathrm{Str}(O^{i_1}\cdots O^{i_n}).
\end{displaymath}
Another object $S$ takes a set of matrices and combines them
(symmetrically) into one matrix:
\begin{equation}
    S^{a_1b_1}_{a_2b_2\ldots a_nb_n} O^{i_2a_2b_2}\cdots O^{i_na_nb_n}
    = \mathrm{Sym}(O^{i_2}\cdots O^{i_n})^{a_1b_1}
\end{equation}

\end{document}